# Ensemble vs. Local Restructuring of Core-shell Nickel-Cobalt Nanoparticles upon Oxidation and Reduction Cycles

Sophie Carenco,[1,*] Cecile S. Bonifacio,[2,^] Judith C. Yang[2,3]

[1] Sorbonne Université, CNRS, Collège de France, Laboratoire de Chimie de la Matière Condensée de Paris, 4 Place Jussieu, 75252 Paris, France.

[2] Department of Chemical and Petroleum Engineering, University of Pittsburgh, 4200 Fifth Avenue, Pittsburgh, Pennsylvania 15260, United States

[3] Department of Physics, University of Pittsburgh, 4200 Fifth Avenue, Pittsburgh, Pennsylvania 15260, United States

[^] E.A. Fischione Instruments Inc., 9003 Corporate Circle, Export, PA 15632, United States (Current affiliation)

* Corresponding author. E-mail: sophie.carenco@sorbonne-universite.fr

## Abstract

Bimetallic nanoparticles are widely studied, for example in catalysis. However, possible restructuring in the environment of use, such as segregation or alloying, may occur. Taken individually, state-of-the-art analytical tools fail to give an overall picture of these transformations. This study combines an ensemble analysis (near-ambient-pressure X-ray photoelectron spectroscopy) with a local analysis (environmental transmission electron microscopy) to provide an *in situ* description of the restructuring of core-shell nickel-cobalt nanoparticles exposed to cycles of reduction and oxidation. It reveals a partial surface alloying accompanied with fragmentation of the shell into smaller clusters, which is not reversible. Beyond this case study, the methodology proposed here should be applicable in a broad range of studies dealing with the reactivity of mono- or bi-metallic metal nanoparticles.

## 1. Introduction

Understanding the behavior of bimetallic nanoparticles exposed to reactive gas is essential to the optimization of their structure and composition in several fields, such as catalysis.[1],[2]

Near-ambient-pressure x-ray photoelectron spectroscopy (NAP-XPS) allows the analysis of surfaces exposed to a few mbar of gas.[3],[4] In the last decade, several studies demonstrated its relevance for the analysis of bimetallic nanoparticles prepared from colloidal synthesis. In core-shell RhPd nanoparticles, metal atoms were shown to reversibly migrate from core to surface as a function of the gas atmosphere (NO, $O_2$, CO, $H_2$ and mixtures).[5] Such transformation was shown not to be perfectly reversible, using a comparison with RhPd single crystal.[6] Other core-shell and alloyed nanoparticles were studied, eg. PtPd and RhPt,[5],[7] RhPd with various Rh:Pd ratio,[8] and AuPd.[9] Several studies were also conducted on bimetallic nanoparticles prone to form metal oxides: CoPt,[10,11],[12] CuNi,[13] CuCo, [13],[14] FeM (M=Pt, Au, Rh),[15] etc. In the case of core-shell nanoparticles, quantitative analysis of XPS depth profiling provides the core and shell thickness, as exemplified with M-$Fe_2O_3$ core-shell nanoparticles (M = Au, Pt, Rh).[16] As an ensemble measurement, NAP-XPS is relevant when the individual nanoparticles do not undergo strong changes in their shape and volume.[17] In order to verify this, other NAP-XPS studies of nanoparticles exposed to gas were backed-up by *ex situ* transmission electron microscopy (TEM).[7,11,15,16],[14,18] In this purpose, the nanoparticles were treated under conditions (gas pressure, temperature) comparable with these of NAP-XPS and analyzed afterwards.

This methodology is efficient but has its limitations when it comes to in-depth analysis of the morphology and structural evolution of the nanoparticles. First, the nanoparticles are exposed to air between the treatment and the TEM observation: this may change the oxidation state of the surface in the case of non-noble metal nanoparticles, sensitive to oxidation. Second, the nanoparticles are cooled down to room temperature and taken out of the reactive atmosphere prior to observation: the structure observed may be a relaxed one instead of that existing under reactive conditions.

Environmental transmission electron microscopy (ETEM), as a local *in situ* analysis technique, provides the opportunity to study the nanoparticles under reactive atmosphere, in a pressure range that is similar to that of NAP-XPS.[19],[20] For example, it was used to follow the formation of NiGa alloy nanoparticles and there deactivation during catalytic methanol formation.[21,22] *In situ* imaging with simultaneous spectroscopy during reactions in the ETEM is feasible providing direct evidence of structural and elemental changes within the bimetallic nanoparticles during reactions.[23],[24] However, a general limitation of this technique is its local character: only a few nanoparticles may be analyzed in-depth.

Core-shell nanoparticles are of interest in model studies that investigate the behavior of metals within confined domains.[25] In this context, nickel-cobalt bimetallic nanoparticles were selected because they are made of abundant metals and present both magnetic and catalytic properties. At 350 °C and in the presence of $H_2$, these nanoparticles reduce $CO_2$ in CO and methanol, instead of forming methane, as would be expected from a cobalt surface.[18] Migration of nickel from the core to the surface was evidenced during the pretreatment step of the nanoparticles, that is, during the cycles of oxidation and reduction that are typically used to clean the nanoparticles surface in the temperature range 220-270 °C.[18] Moreover, thermal properties of these nanoparticles were investigated by TEM, showing that the core-shell structure was stable up to 440 °C in the absence of reactive gas.[26]

In the present paper, we combined NAP-XPS and ETEM to provide a precise account of the structural and chemical transformation occurring in core-shell nickel-cobalt nanoparticles under cycles of oxidation and reduction in the mbar pressure range. An unprecedented description of the surface oxidation state during cycles of oxidation and reduction in the mbar pressure regime is provided. At the ensemble scale, NAP-XPS demonstrates that, if nickel can be fully reduced under $H_2$, this is not the case for cobalt whose average oxidation does not goes bellow oxidation state (+I). At the local scale, ETEM confirmed this finding, from both structural (selected area diffraction) and spectroscopic (electron energy loss spectroscopy) data. Moreover, TEM imaging and elemental mapping evidenced size evolution and reconstruction of the core-shell structure of the nanoparticles as a result of the gas treatments. These findings from combined ensemble and local *in situ* analyses dispute the ideal picture of reversible restructuring of bimetallic nanoparticles when switching between oxidizing and reducing conditions.

## 2. Experimental Section

The Ni-Co core-shell nanoparticles with

sizes ranging from 25 to 45 nm were synthesized in a two-step process.[18] Ni-Co core-shell NPs were well dispersed in hexane using an ultrasonic bath for 2 minutes.

NAP-XPS measurements were conducted at beamline 11.0.2 in the Advanced Light Source in Berkeley, California.[27] Two cycles of oxidation (*I-O2*, *III-O2*) and reduction (*II-H2*, *IV-H2*) at respectively 1 Torr at 220 °C and 5 Torr at 270 °C were performed (Figure 1). A sub-monolayer of nanoparticles was drop-casted on a gold foil, and gold could still be observed in the Au 4f region, which served for calibration, and Au 5p region, which overlapped with the Co 3p region. Each XP spectrum was collected within a few minutes with the photon energy of 250 eV, 500 eV, 700 eV and 875 eV, tuned to collect electrons with various mean free paths, evaluated for the Ni 3p – Co 3p region:[28],[29],[30] 0.5 nm, 0.8 nm 1.1 nm and 1.3 nm, respectively. Resistive heating of the sample-holder button heater was used to adjust the temperature to the target.

Spectrum processing was carried out using the CasaXPS software package. Shirley background was used for all spectra.

ETEM studies were performed using a dedicated ETEM FEI Titan (Center for Functional Nanomaterials, Brookhaven National Laboratory) operated at 300 keV. Similar to the XPS experiments, two cycles of oxidation-reduction reactions were conducted at 0.2 Torr at 220 °C and 0.3 Torr at 270 °C. Longer reaction times in the ETEM were used (Figure 1) to compensate for the lower allowable ETEM reaction pressures in comparison to the NAP-XPS reaction pressures (0.2–0.5 Torr in the ETEM *vs.* 1–5 Torr in the NAP-XPS). Electron energy loss spectroscopy (EELS), dark field (DF) Scanning TEM (STEM) images and selected area electron diffraction (SAED) were recorded during oxidation or reduction reactions in the ETEM.

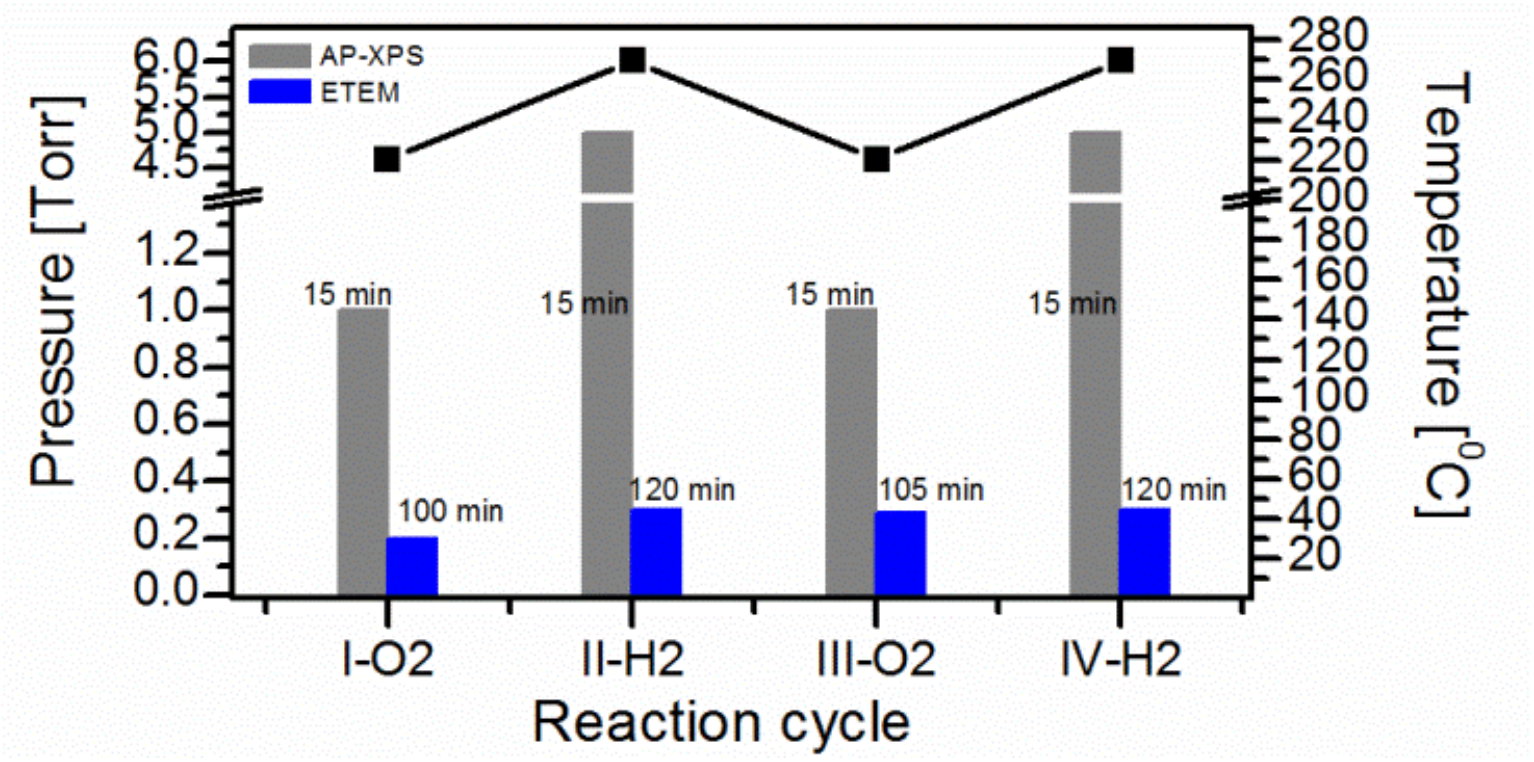

Figure 1: The four stages of treatment (*I* to *IV*) in NAP-XPS and ETEM. Pressures are indicated in grey for NAP-XPS and in blue for ETEM.

## 3. Results and Discussion

### 3.1 Ensemble measurement by NAP-XPS

In order to analyze of the evolution of oxidation state of Ni and Co, a sub-monolayer of nanoparticles was drop-casted on a gold foil, and a set of NAP-XPS data was collected at four different photon energy (250, 500, 700 and 875 eV).

Figure 2 shows the spectra collected on the region of Ni 3p and Co 3p. Au 5p3/2 was also observed in this region. The bottom spectrum corresponds to the initial state of the nanoparticles, after introduction in the XPS chamber, while the spectra *I* to *IV* were collected after each step of oxidation or reduction. Detailed fitting procedure can be found in ESI (Table S1). For the sake of clarity, all components were plotted, included $3p_{1/2}$ components and satellites, but only those discussed below were pointed out with colored vertical lines.

*Initial State.* After introduction in the XPS chamber, the signal-to-noise ratio was fairly low due to the presence of carbonated ligands on the surface of the nanoparticles (Figure 2, bottom spectrum). Cobalt could be observed in an oxidized form, noted $CoO_x$, with a peak of binding energy (B.E.) 60.2 eV (orange dotted line).

*First oxidation treatment (*I-O2*).* The first oxidation treatment allowed burning the ligands and recovering more signal (Figure 2, spectrum *II*). As expected, both cobalt and nickel were observed in an oxidized form, with peaks at 60.2 eV (orange) and 66.9 eV (light green), respectively. With a photon energy of 700 eV, corresponding to an IMFP of ca 1.1 nm, the Ni:Co ratio was of 1:5 (top-right of Figure 2, *I-O2*).

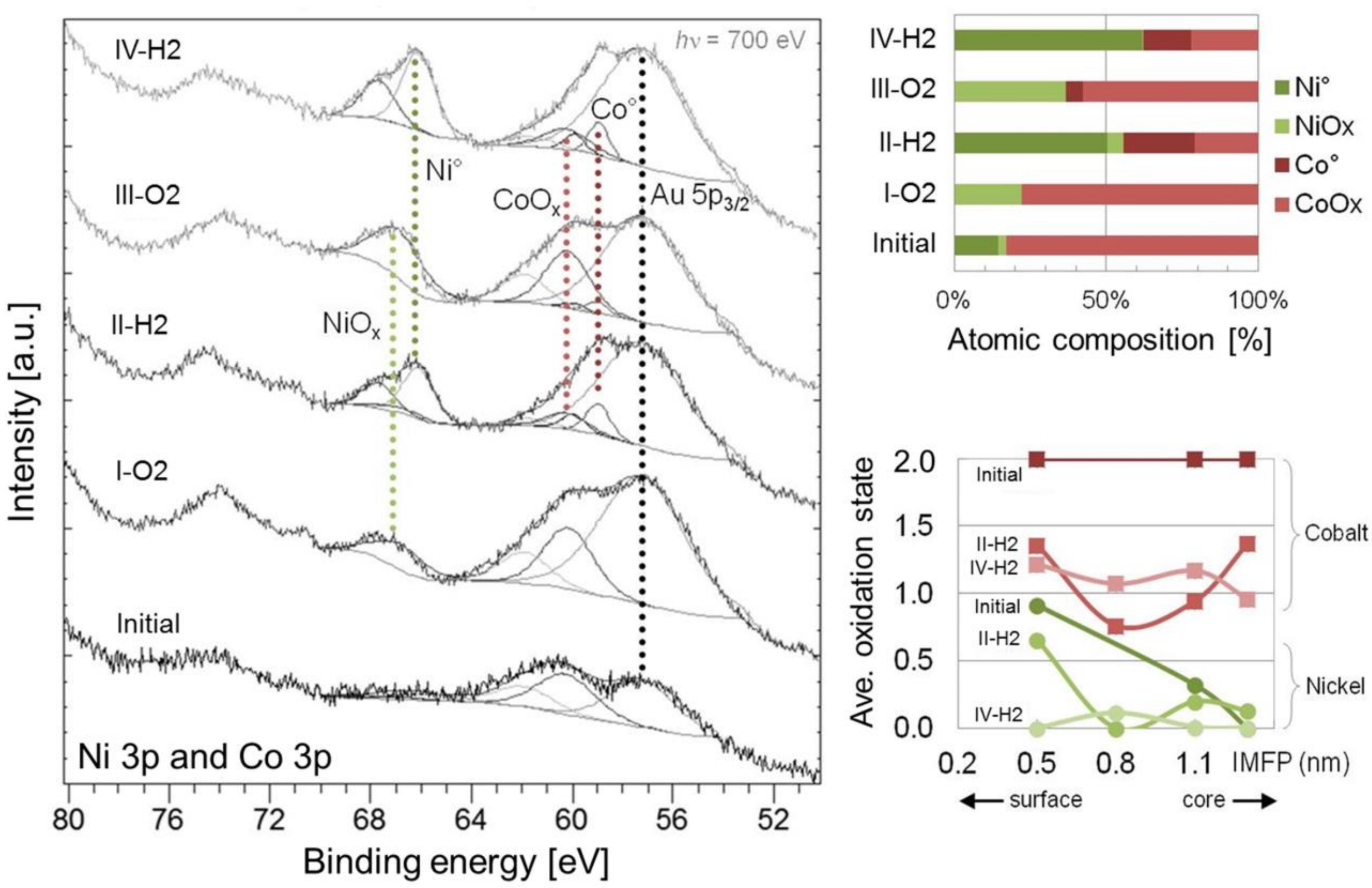

Figure 2: NAP-XPS of NiCo nanoparticles during cycles of oxidation and reduction. (Left) Photoelectron spectra of Ni 3p and Co 3p region collected with a photon energy of 700 eV. Dotted line are a guide to the eye: Au $5p_{3/2}$ (black), metallic Ni $3p_{3/2}$ (dark green), NiO (light green), metallic Co $3p_{3/2}$ (brown) and $CoO_x$ (orange). (Top right) Atomic composition in %, extracted from the spectra. (Top bottom) Average oxidation state of Ni and Co extracted from spectra collected with four different photon energies.

Cobalt oxides are a mixture of CoO and $Co_3O_4$, which are more easily discriminated using the Co 2p region (Figure S2-E in the ESI). This requires the surface to be analyzed with more energetic photons of 1115 eV and allows to discriminate CoO (made of Co(II) only) from $Co_3O_4$ (mixed valence state). We observed that CoO was the predominant oxide after the oxidation treatment. Overall, only 20 % of the cobalt was in the $Co_3O_4$ environment. Based on these Co 2p data, the average oxidation state of cobalt was 2.07 (Figure S2-C).

*First reduction treatment (*II-H2*).* Upon reduction, new XPS peaks appeared, indicating the formation of Co(0) and Ni(0), at 66.0 eV (dark green) and 58.9 eV (brown), respectively (Figure 2, *II-H2*). While the nickel was almost completely reduced by the treatment, cobalt oxide still represented about half of the total cobalt (Figure 2 top-right, *II-H2*). This was confirmed on the Near Edge X-Ray Absorption Fine Structure (NEXAFS) spectrum at Co L-edge (Figure S3 in the ESI), that shows features from both Co (778 eV) and CoO (776.5 and 781 eV).[31,32] Analysis of the Co 2p region confirmed the remaining of CoO (ca 20 % of all Co) and allowed to detect a smaller fraction of $Co_3O_4$ (Figure S2-B).

Additionally, the Ni:Co ratio significantly increased to ca 1:1, confirming that nickel was able to migrate from the nanoparticles core to the near-surface, as a consequence as the reduction treatment (Figure 2 top-right, *II-H2*), and consistently with preliminary results reported in a previous study.[18]

*Second oxidation treatment (*III-O2*).* The second oxidation treatment only slightly reduced the Ni:Co ratio down to about 2:3, while both nickel and cobalt underwent oxidation (Figure 2 top-right, *III-O2*). Oxidation states were similar to those of step *I-O2*. This was confirmed by analysis of the Co 2p region (Figure S2-B). NEXAFS at Co L-edge also indicated the presence of both CoO species (identified by the shoulder at 776.5 eV) and $Co_3O_4$ species (identified by the peak at 779.6 eV), although both strongly overlap.

*Second reduction treatment (*IV-H2*).* A last reduction treatment increased the Ni:Co surface ratio to ca 3:5. The nickel could be fully reduced to a metallic state while the cobalt remained partially oxidized, as observed at the first reduction step (Figure 2 top-right, *IV-H2*).

Overall, relative ratio of nickel and cobalt showed that nickel was able to diffuse from the core to the shell, and the surface was slightly richer in nickel under reducing conditions.

The average oxidation states of nickel and cobalt were plotted as a function of IMFP (Figure 2 bottom-right). From the 3p region, average oxidation state was calculated assuming Ni(II) and Co(II) states for the oxides, and using the relative amounts of reduced and oxidized species showed on the top graph. For the sake of clarity, the steps *I* and *III*, where both metals are fully oxidized, were omitted.

Concerning nickel, the analysis at the initial step showed that, deeper into the core, the nickel is more reduced. This made sense, as in this case the oxidation comes from the exposure of the nanoparticles to the air during the washing treatment, at the end of the colloidal synthesis: at room temperature in air, only the very surface gets oxidized. For step (II), only nickel closer to the surface was found to be significantly oxidized. For step (IV), the nickel was metallic at all depths probed. Regarding cobalt, the situation was more complex, as purely metallic cobalt was never observed. At initial step, the whole thickness of the cobalt shell was fully oxidized, which made sense since even the nickel right beneath it was slightly oxidized. For steps (II) and (IV), no clear trend was observed in terms of depth profile, also similar average oxidation states in between 0.8 and 1.4 were observed.

Knowing that, at step (IV), the surface of the nanoparticles contains as much nickel as cobalt, one could postulate that the reduced cobalt atoms are closer to the reduced nickel, while the oxidized cobalt atoms could form islands on the particles.

Altogether, this experiment revealed a complex speciation of nickel and cobalt at oxidation state between 0 and II (and even III for cobalt). The near surface region was globally more oxidized than the inner shells. Nickel was fully reduced by $H_2$ and partly migrated toward the outer shells, while cobalt was only partially reduced.

In order to assess the localization of the elements within the nanoparticles as well as their morphological changes, ETEM was performed with the same sequence of oxidations and reductions.

## 3.2 Local measurement in ETEM

*Morphology and structure*

Dark-field (DF) STEM images (Figure 3-top) show the morphological changes in the nanoparticles during the two cycles of oxidation-reduction reactions. During oxidation cycle *I-O2*, the surface of the nanoparticles was covered with surface oxide of both nickel oxide and cobalt oxide. Interestingly, in the low pressure range used here, no hollowing of the particles was observed as a consequence of oxidation. Rather, slight volume increase was detected during the *I-O2* step (Figure S4). Using image processing methods, the DF images from the oxidation and reduction steps were colorized as red and green, respectively. By overlaying the images from *I-O2 and II-H2* cycles, a decrease in the overall volume of the particles was demonstrated, as a result of the reduction (Figure S4). This was accompanied by a partial fragmentation of the nanoparticles on their outer regions, which released smaller objects, observed on Figure 3 and S4 (pointed by arrows). The larger nanoparticles were mostly crystalline, as showed by SAED (Figure 3 bottom). There was no significant change in volume observed with further oxidation and reduction of the nanoparticles *during III-O2* and *IV-H2*, respectively.

In terms of mechanical behavior, these observations suggested a dense and cohesive core surrounded by a more brittle shell, which accommodated for most of the volume changes associated during the first oxidation (expansion) and reduction (shrinking) steps. In order to correlate this to XPS data, composition and oxidation state of the outer region of the nanoparticles were analyzed in greater details.

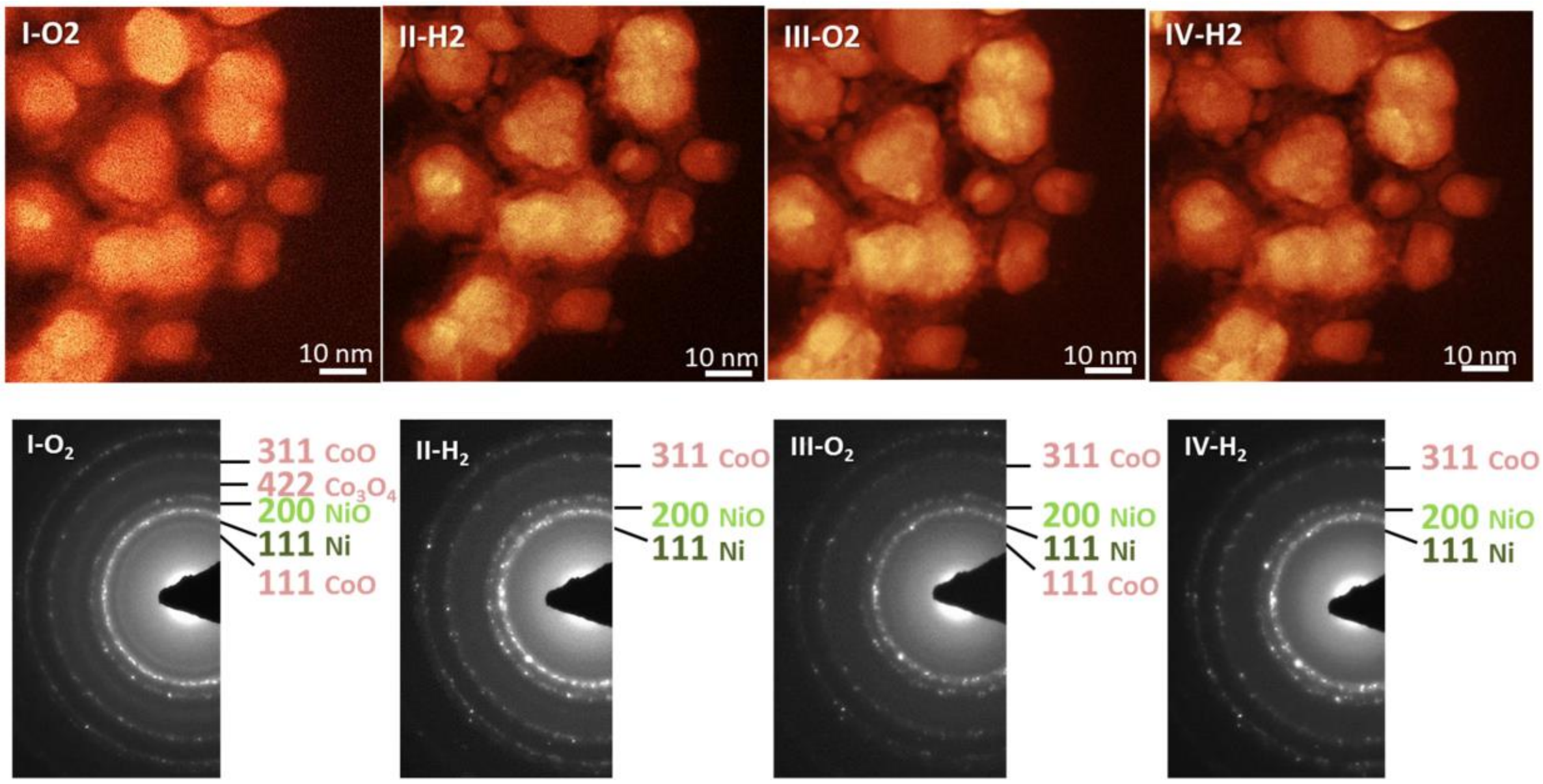

Figure 3: (Top) STEM-HAADF from step (I) to step (IV). (Bottom) SAED of the corresponding regions.

*Elemental mapping from EELS*

The local elemental composition of the nanoparticles was analyzed by acquiring EELS spectrum images of the Co L edge and Ni L edge. In the initial state and during the cycles of oxidation and reduction, the nickel-rich core and the cobalt-rich shell were observed.

The EELS spectrum images provided the distribution of Ni and Co during the reactions. During the first reduction step (*II-H2*), clusters of particles were observed on the surface of the nanoparticles (Figure 4). This correlates with the STEM image and SAED analysis during *II-H2* in Figure 3. Further analysis of the EELS maps identified the clusters as Co-rich (Figure S5). The spectrum map during *IV-H2* step in Figure 4 shows green pixels, representing Ni, on the surface of the nanoparticle, suggesting rearrangement of the core-shell structure.

EELS analysis of the composition in the near-surface is presented on the right panel of Figure 4. The oxygen ratio is always fairly high (around 50 %) as it comes not only from the nanoparticle surface oxide, but also for the underlying grid. It will not be further discussed. As expected, on the shell region nickel was detected in much lower amounts than cobalt. However, the atomic fraction of nickel increased to 4.5 %, 4.9 % and 12.0 % from the initial state to the *II-H2*, the *III-O2* and the *IV-H2* steps, respectively. Further analysis of the spectrum maps showed an increased fraction of nickel in the shell from *II-H2* to *IV-H2* (Figure S6).

These observations confirmed the partial migration of Ni from the core to the shell. They also indicated that the evolution of Ni:Co ratio observed by XPS (Figure 2) were mostly a surface phenomenon, and that full alloying of the nanoparticles was not occurring under these fairly soft reaction conditions.

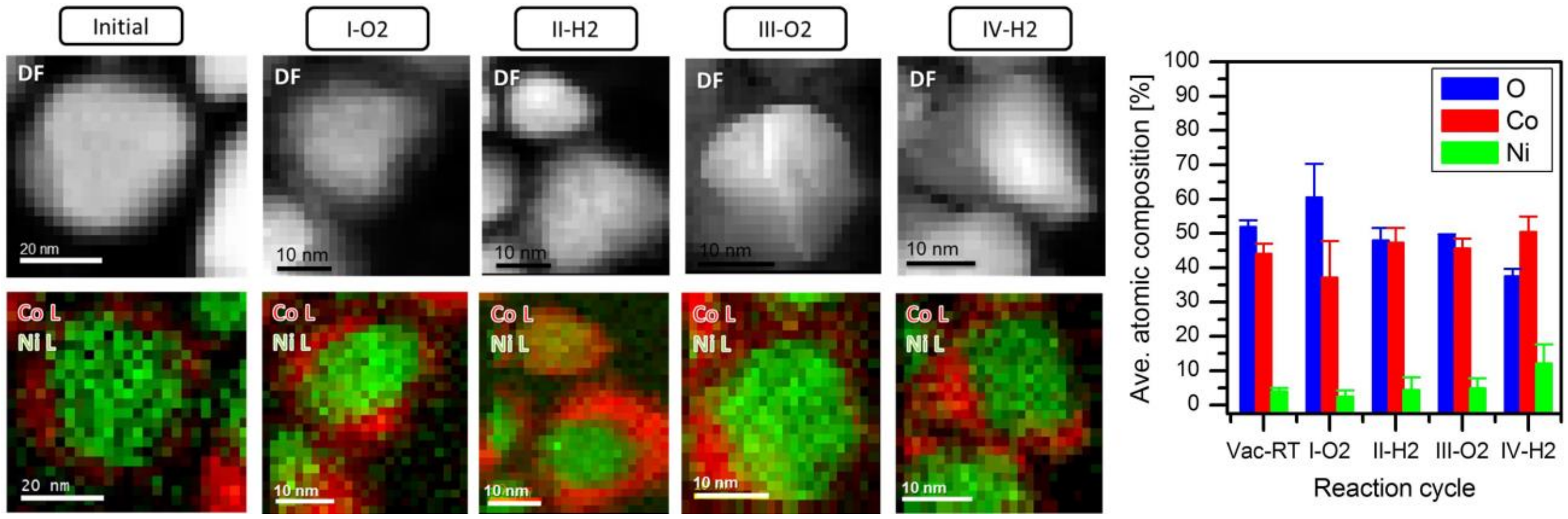

Figure 4: Dark field STEM images (top) and corresponding spectrum images for Ni L edge and Co L edge (bottom) at each stage. Initial state under UHV, (*I-O2*) After oxidation by 1 Torr $O_2$ at 220 °C, (*II-H2*) After reduction by 5 Torr $H_2$ at 320 °C, (*III-O2*) After second oxidation by 1 Torr $O_2$ at 220 °C, (*IV-H2*) After second reduction by 5 Torr $H_2$ at 320 °C. Cobalt is in red and nickel in green. (Right) Near-surface composition of the nanoparticles at each stage (within 4 nm of the surface) extracted from the EELS spectrum images.

*Oxidation State*

Oxidation state was evaluated by two methods: through the identification of crystalline phases by SAED analysis and through the Co $L_3/L_2$ ratio measured by EELS.

Figure 3 shows the changes in the nanoparticles along with the SAED. As a first result, SAED showed that metallic nickel was present at all stages in the nanoparticles. Because it was not always observed by XPS (a surface-sensitive spectroscopy), it had to be mostly in the core of the nanoparticles. Moreover, metallic cobalt was never observed. We proposed this is a consequence of both the lesser amount and smaller size of the corresponding domains on the surface, yielding broader and weaker SAED spots, which were thus not detected in these measurements.

During oxidation cycle *I-O2*, the surface of the nanoparticles was covered with both nickel oxide and cobalt oxide. SAED analysis resulted to identification of the oxide species as NiO, CoO and $Co_3O_4$ (*I-O2* in Figure 3b). These observations were consistent with XPS and NEXAFS results.

After reduction cycle *II-H2*, the surface of the nanoparticles was covered with clusters of particles (12-17 nm in size). SAED spots for $Co_3O_4$ were absent, as well as the weaker ones from CoO (111). This confirms the partial reduction of cobalt oxides, observed by XPS. This process was essentially repeated upon steps (III) and (IV) even though $Co_3O_4$ was not clearly observed anymore. This was once more consistent with XPS, which showed the coexistence of Co(0) and CoO at step (III), the cobalt being overall slightly more reduced than at step (I).

Complementarily with the SAED study, the oxidation states of cobalt during the reaction cycles were determined from the $L_3/L_2$ intensity ratio calculated from the acquired Co $L_{2,3}$ ionization edges (Figure S7). The average EEL spectra at the shell was obtained and the branching ratios calculated using the method described by Zhao et al.[33] where 6 Gaussian peaks (3 for $L_3$ and 3 for $L_2$ edges, see Figure S8 for details of analysis) were used to fit the $L_3$ and 3 for $L_2$ edges. The $L_3/L_2$ intensity ratio was subsequently calculated from the ratio of the integrated areas underneath the Co $L_3$ and $L_2$ absorption edges, respectively. To identify the phase of the cobalt oxide, EELS data were acquired from of $Co_3O_4$ and CoO powders and the same fitting method was applied. In the case of metallic Co, the metallic Co EELS data from Zhao et al. was analyzed and used as reference for the metallic Co phase. The calculated $L_3/L_2$ intensity ratios were compared in Figure 5 (right) to identify the phase of the cobalt oxide during each reaction cycles in the ETEM study. The Co $L_3/L_2$ ratio was decreasing for the reference spectra for CoO, Co and $Co_3O_4$, respectively. By comparing the experimental results to the reference spectra, the surface of the nanoparticle has undergone a change in valence states between $Co^{2+}$ and $Co^0$ during the oxidation-reduction reactions in the ETEM. The decreasing and increasing $L_3/L_2$ ratio exemplified the reduction and oxidation cycles, respectively.

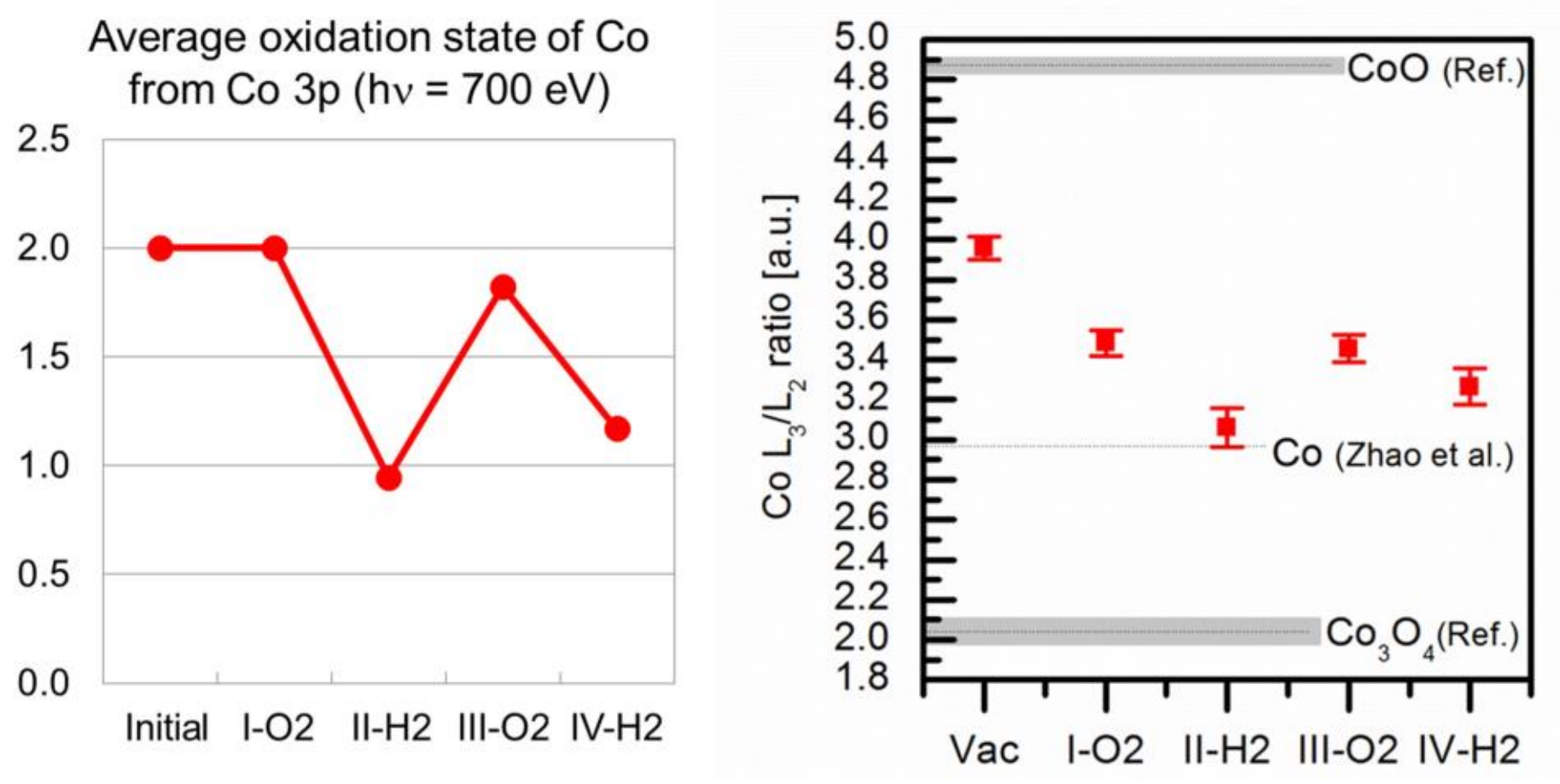

Figure 5: Oxidation state of cobalt at the surface of the nanoparticles during the treatments. Left: from Co 3p with a photon energy of 700 eV. Right: from Co $L_3/L_2$ ratio measured by EELS.

During the *I-O2* step, the $L_3/L_2$ ratio is expected to be equal or higher than the $L_3/L_2$ ratio at room temperature and under vacuum. Such discrepancy can be attributed to residual oxygen inside the microscope before the reaction cycle and possibly from data analysis. The accuracy of $L_3/L_2$ EELS intensity ratios is subject to processing parameters during data analysis such as continuum background fitting, widths of integration windows, etc.[34] As a result, variation of the absolute $L_3/L_2$ intensity ratios from the reference sample is observed. However, the relative changes observed in this study are consistent and impart reliable evidence for the variation in 3d transition metal oxidation states.[35],[36] In fact, the convexity of the Co $L_3/L_2$ ratio correlated with valence state of Co in the XPS results from *II-H2* to *IV-H2* steps in Figure 5 (left). In both XPS and EELS results, the CoO phase is present during the oxidation-reduction steps, in contrast with the $Co_3O_4$ phase. This may be explained by the preferential formation of CoO at the nanometric scale compared to the bulk.[37],[38]

## 4. Conclusion and outlooks

The unique combination of ensemble measurement (NAP-XPS with depth profile, NEXAFS) with local measurements (ETEM, DF-STEM, SAED and EELS) allows drawing a full and detailed picture of the behavior of Ni-Co core-shell nanoparticles exposed to oxidation and reduction cycles.

(1) the nickel-rich metallic core of the nanoparticles was mostly not affected by this procedure, as showed by its chemical and structural stability.

(2) the surface of the nanoparticles was mechanically brittle, possibly due to volume increase upon oxidation. Partial fragmentation of the surface layers was not a reversible process.

(3) both techniques confirm the trend of nickel enrichment as a consequence of the cycles, although there is a slight pullback upon re-oxidation at step (III). This pullback may tentatively be attributed to the lower Gibbs free energy of CoO vs. NiO.[23]

(4) both techniques confirmed that reduction of cobalt was only partial at steps (II) and (IV), and slightly more advanced at the later one.

At this stage, there is no indication regarding further morphological and compositional evolution of the nanoparticles upon further cycling. However, the lack of reversibility observed within the first two cycles suggests that deeper changes may occur over time ad cycling in a real setup. Increase of overall free energy (from surface oxidation and mixing of metals) should be the driving force for these irreversible changes.

With the combined XPS and TEM studies, the ensemble and local elemental composition and oxidation states were obtained, allowing for an understanding of the stability limitations of the Ni-Co core-shell NP structure. None of the two techniques, albeit multifaceted, may have provided a satisfactory picture of the overall transformation of the nanoparticles, hence on the nature of active sites. Moreover, all the transformations described above occurred under fairly gentle conditions (below 280 °C and 5 Torr), compared with these typically used in hydrogenation processes, eg. CO and $CO_2$ hydrogenations, or in oxidation catalysis. Thanks to the growing number of ETEM and NAP-XPS instruments worldwide, we believe that the methodology presented in this article should find an application in a broad range of studies dealing with the reactivity of mono- or bi-metallic metal nanoparticles.

## Acknowledgements

The *Advanced Light Source* is acknowledged for beamtime on beamline 11.0.2. Hendrik Bluhm is acknowledged for his expertise with NAP-XPS at beamline 11.0.2. S.C. acknowledges CNRS, Sorbonne Université and Collège de France. C.S.B. and J.Y. acknowledge Sara Mills for the help in EELS data processing, Dr. Eric Stach and Dr. Huolin Xin from Center for Functional Nanomaterials (CFN) at Brookhaven National Lab for the ETEM data acquisition. The ETEM work was done in CFN, BNL supported by the Office of Basic Energy Sciences of the US Department of Energy Contract No. DE-SC0012704.

## Table of Content Graphic

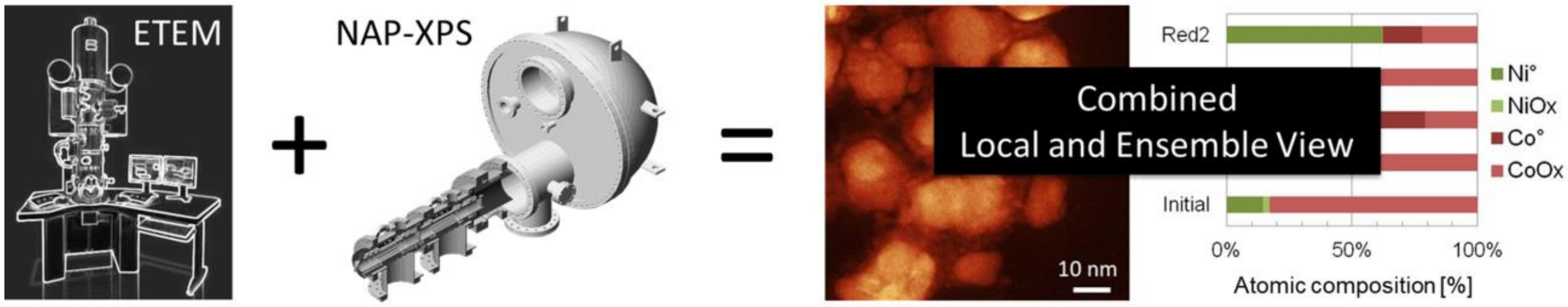

Bring colour and life to core-shell nanoparticles! Combining local and ensemble environmental analyses provides unique insight on restructuring under reactive gases.